\documentclass[journal=jacsat,manuscript=article]{achemso}

\usepackage[version=3]{mhchem} 
\usepackage{amssymb}
\usepackage{amsmath}
\usepackage{gensymb}
\usepackage{graphicx}
\usepackage{float}
\usepackage{gensymb}
\usepackage{textgreek}

\author{Arvind Shankar Kumar}
\author{Mingyuan Wang}
\author{Yancheng Li}

\affiliation{
Department of Physics, Case Western Reserve University, 2076 Adelbert Road, Cleveland, Ohio 44106, USA
}
\author{Ryuji Fujita}
\affiliation{
Department of Physics, Oxford University, Parks Road, Oxford, OX1 3PU, United Kingdom
}
\author{Xuan P.A. Gao}
\affiliation{
Department of Physics, Case Western Reserve University, 2076 Adelbert Road, Cleveland, Ohio 44106, USA
}
\email{xuan.gao@case.edu}

\title[]
  {Interfacial charge transfer and gate induced hysteresis in monochalcogenide InSe/GaSe heterostructures}

\abbreviations{IR,NMR,UV}
\keywords{American Chemical Society, \LaTeX}

\begin{document}


\begin{abstract}
 Heterostructures of 2D van der Waals semiconductor materials offer a diverse playground for exploring fundamental physics and potential device applications. In InSe/GaSe heterostructures formed by sequential mechanical exfoliation and stacking of 2D monochalcogenides InSe and GaSe, we observe charge transfer between InSe and GaSe due to the 2D van der Waals interface formation and a strong hysteresis effect in the electron transport through the InSe layer when a gate voltage is applied through the GaSe layer. A gate voltage dependant conductance decay rate is also observed.  We relate these observations to the gate voltage dependant dynamical charge transfer between InSe and GaSe layers.
\end{abstract}

\section{Introduction}
Heterostructures built from two-dimensional (2D) van der Waals (vdW) layered materials offer a large variety of possible systems for probing fundamental physics and device applications, thanks to the development of facile mechanical exfoliation and dry transfer techniques\cite{ Novoselov2016,Geim2013, Frisenda2018}. Using single crystal hexagonal boron nitride (hBN), a 2D vdW insulator as the substrate, graphene/hBN and MoS$_2$/hBN heterostructures exhibited remarkably improved electron transport performance compared to bare graphene or MoS$_2$ on amorphous oxide dielectric \cite{Dean2010,Dean2013,Cui2015}. In addition to the high quality interface and carrier mobility, the coupling between the two vdW crystals' lattices in a heterostructure may also lead to new phenomena. The richness of emerging physics in 2D vdW heterostructures is exemplified by the plethora of novel quantum transport phenomena discovered in graphene/hBN heterostructures in recent years such as the fractal quantum Hall effect, topological conduction, and unconventional superconductivity\cite{Dean2013,Terres2016,Li2016,Cao2018}.

Heterostructures built from 2D vdW semiconductors have also demonstrated merits of usage as optoelectronic devices \cite{Zhang2016,Liu2016}. A lot of the research in this direction has been exploring MoS\textsubscript{2} or other transition metal dichalcogenide (TMDC) based heterostructures \cite{Lee2014,Furchi2014,Cheng2014,Deng2014,Ye2016,Xue2016,Liu2016,Yu2013,Wang2015,Baugher2014}. For photodetector applications, graphene/TMDC/graphene heterostructures and p-n junction type vertical TMDC heterostructures have emerged as two leading candidates \cite{Liu2016}. In graphene/TMDC/graphene photodetectors, the TMDC layer is used as the primary light absorbing layer and graphene is used as a contact material whose gate tunable work function is used to drive the generated charge carriers to opposite electrodes. In this category, graphene/MoS\textsubscript{2}/graphene heterostructures have shown highly effeicient photocurrent generation with a maximum External Quantum Efficiency (EQE) of 55 percent \cite{Yu2013}. In the case of p-n junction type vertical vdW heterostructure photodetectors, the main advantage over conventional p-n junction photodiodes is the possibility of modifying diode characteristics and photoresponsivity through gate modulation \cite{Liu2016}. In this category, Vertically stacked MoS\textsubscript{2}/WSe\textsubscript{2} \cite{Lee2014, Cheng2014},MoS\textsubscript{2}/Black Phosphorous \cite{Deng2014} and MoS\textsubscript{2}/GaTe heterostructures \cite{Wang2015} have been explored as photodetectors, with EQE as high as 61 percent reported in MoS\textsubscript{2}/GaTe heterostructures \cite{Wang2015}. This type of heterostructure has also been explored for light-emitting devices\cite{Baugher2014}, encouraged by the observation of electroluminescence in MoS\textsubscript{2}/WSe\textsubscript{2} heterostructures\cite{Cheng2014}. 

\begin{figure*}[]
\includegraphics[width=16 cm]{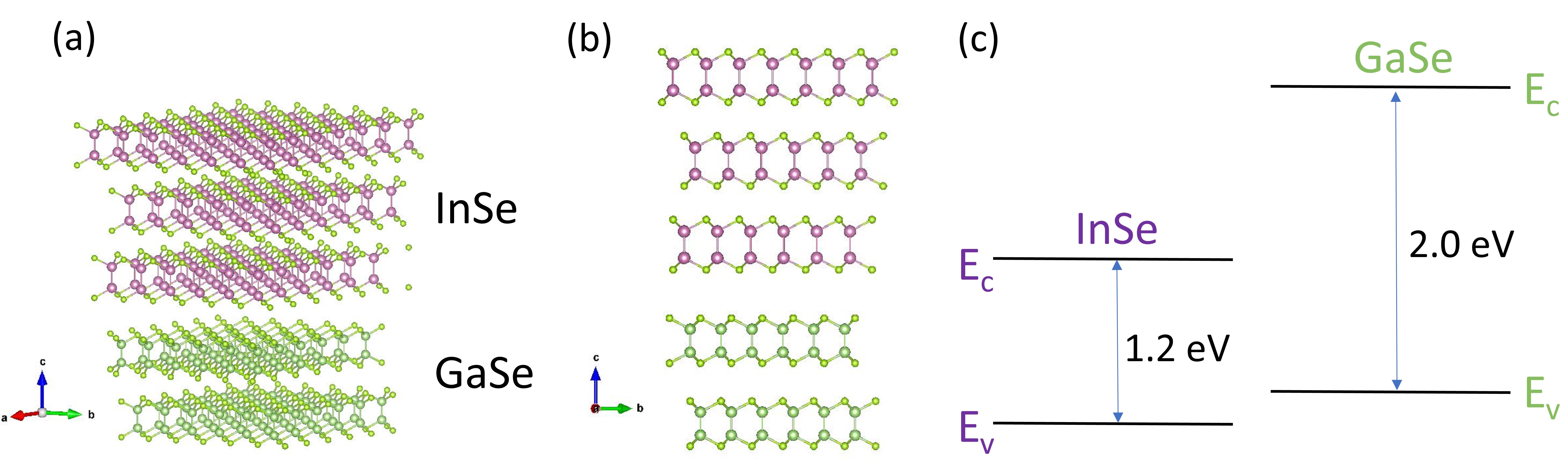}
\caption{\label{fig:epsart} Crystal structure. (a),(b) InSe/GaSe crystal structure. (c) Type II staggered band alignment in InSe/GaSe heterostructures. }
\label{crystal details}
\end{figure*}

Going beyond MoS\textsubscript{2} and TMDCs,  group-III 2D monochalcogenides, represented by InSe, and GaSe, have emerged as an important family of 2D semiconductors with unique attributes. InSe has a direct band gap of 1.2 eV in the bulk and it has a layered honey-comb lattice structure with each layer containing four covalently bonded atomic planes, given by the bonding structure of Se-In-In-Se\cite{Balakrishnan2014,Feng2014,Sucharitakul2015,Bandurin2017}. The order of layer stacking gives us different polymorphs – $\beta$-InSe,$\gamma$-InSe etc. GaSe, another group III monochalcogenide 2D semiconductor, has the same crystal structure of InSe but with a wider band-gap of 2 eV, in the bulk form (Fig.1)\cite{Balakrishnan2014}. The high intrinsic electron mobility of InSe has recently attracted significant research interest for exploring fundamental physics such as high mobility 2D electron transport, quantum hall effect \cite{Feng2014,Sucharitakul2015,Bandurin2017,Ho2017,Li2018} and spin-orbit interaction \cite{Premasiri2018, Zeng2018} in an atomically thin semiconductor layer. In terms of optical properties, in contrast to the TMDC family, these III-VI monochalcogenides exhibit a direct to indirect bandgap transition in the few layer thickness limit\cite{Mudd2013,Bandurin2017} suggesting possible device applications in opto-electronics with only a few layer thickness. Few layer InSe nanodevices were demonstrated as flexible photodetectors \cite{Lei2014, Tamalampudi2014} and n-InSe/p-GaSe heterostructures were shown to have potential in light emitting or photodetection devices\cite{Yan2017, Balakrishnan2014}. There is also an increasing interest in the electronic, optical and thermal properties of group-III monochalcogenide heterostructures in theory\cite{Ibarra2017,Rahman2018}. However, the electrical transport properties of InSe/GaSe heterostructures remain largely unexplored.

In this paper, we report on the gate modulated electrical transport properties of InSe/GaSe heterostructures. Evidence for electron transfer from GaSe to InSe upon the heterostructure formation is observed in InSe FET devices supported on SiO\textsubscript{2} after stacking few-layer GaSe on top of InSe. In devices with InSe supported on few-layer GaSe, the gating of InSe channel is shown to exhibit a strong scanning rate dependent hysteresis effect and time-dependent conductance relaxation. Such hysteretic or dynamic phenomenon is explained in the context of type II band alignment in InSe/GaSe\cite{Balakrishnan2014,Yan2017} and gate electric field assisted charge transfer at the interface. These results provide insights on the understanding and future developments of group-III 2D monochalcogenide heterostructure electronic devices.

\section{Experimental Section}
For this study, InSe and GaSe bulk crystals grown by the Bridgeman method, similar to Sucharitakul et al \cite{Sucharitakul2015}. InSe and GaSe nano-flakes were exfoliated using the mechanical exfoliation method\cite{Novoselov2016,Geim2013,Frisenda2018}. First, nanoflakes of GaSe or InSe were transferred onto degenerately doped Si substrates with a 300 nm thick SiO\textsubscript{2} surface capping layer. To form a vertical heterostructure, e.g. InSe/GaSe heterostructure with InSe on top, a PDMS (Poly-DiMethyl-Siloxane) stamp placed on a glass slide was used to pick up an InSe nanoflake of thickness 10 nm – 60 nm and aligned with the GaSe pre-identified on the SiO\textsubscript{2}/Si substrate under a transfer stage microscope, and brought into contact. The PDMS was slowly removed, leaving the formed InSe/GaSe heterostructure on the substrate. GaSe/InSe heterostructure with InSe at the bottom can be formed in a similar way except it is the GaSe nanoflake being manipulated by the PDMS and aligned with InSe nanoflake on SiO\textsubscript{2}/Si substrate. The completed heterostructure was then annealed at 250\degree C under Ar/H\textsubscript{2} atmosphere. To fabricate metal electrodes, a copper grid was used as a shadow mask, with the grid covering the active channel of the heterostructure device. 30 nm thick Ti and 60 nm thick Ni metals were then evaporated onto the masked substrate, followed by removal of the copper grid. The completed device was then annealed at 250\degree C under Ar/H\textsubscript{2} atmosphere to improve contacts with the metal electrodes. 

For electrical measurement at room temperature, devices were characterized in a Lakeshore Probe Station. DC source (S)-drain (D) voltage in the range 0.1-1V and gate voltages from -70V to 70V were applied to the Si substrate backgate (G) and the source-drain current was measured with a current amplifier. A Physical Property Measuring System (Quantum Design Inc.) was used for temperature dependent measurements down to 10K. 

\section{Results and discussion}
\begin{figure*}[]
\includegraphics[width=16cm]{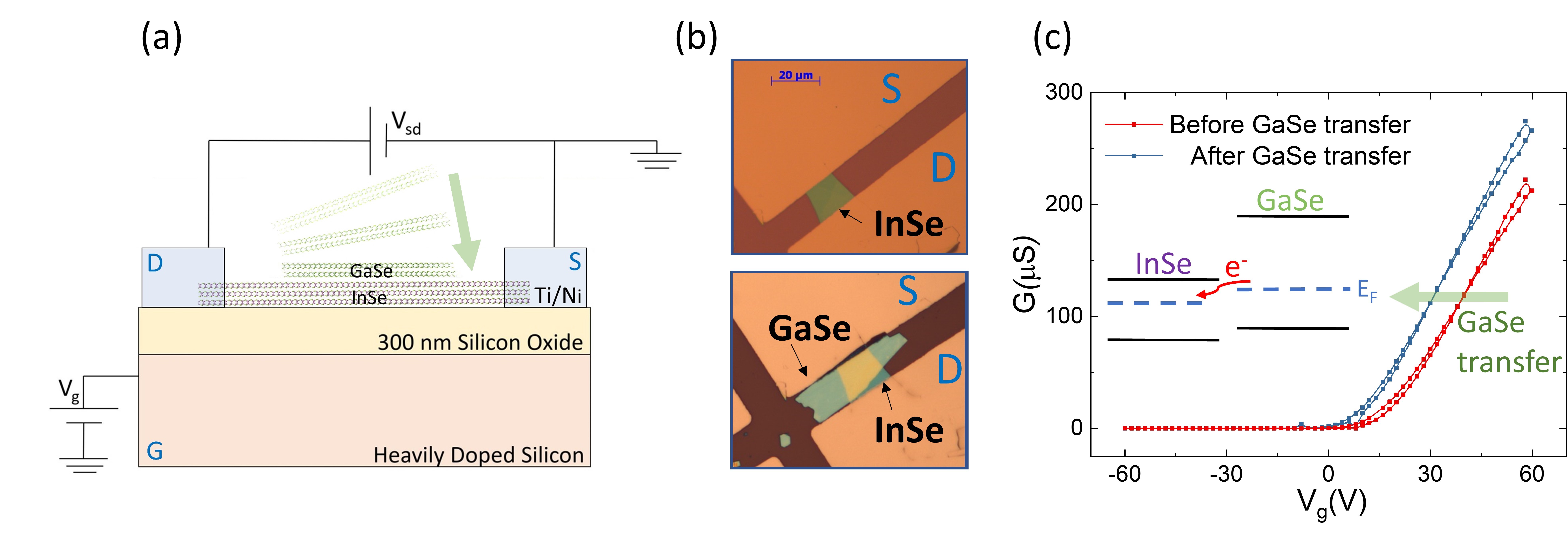}
\caption{\label{fig:epsart}Gate-transfer curves of an InSe nanoflake FET device before and after GaSe nanoflake transfer. (a) Device schematic indicates transfer of GaSe nanoflake onto an InSe nanoflake FET device. Source (S), Drain (D) and Gate (G) terminals are marked on schematic. (b) InSe device optical image before (top) and after (bottom) GaSe transfer.(c) Gate-transfer curves before and after GaSe transfer.}
\label{thresholdshift}
\end{figure*}

 We first consider an experiment where we study the effect of the GaSe/InSe hetero-interface on InSe's electronic property before and after the hetero-interface formation (Fig.\ref{thresholdshift}(a)). In this experiment, a few-layer GaSe flake was placed on top of a two-probe device with an InSe nanoflake conducting channel such that the backgate voltage is applied only through the SiO\textsubscript{2} dielectric and we can use the gate transfer characteristics of InSe before/after the GaSe transfer to understand the influence of GaSe on InSe's transport behavior. For this experiment, we first measured the room temperature gate-transfer characteristics in vacuum of a two-probe InSe FET device on SiO\textsubscript{2} (Fig.\ref{thresholdshift}(b), top image). The measurement was done by setting the source-drain voltage $V_{sd}$ (at 0.5V) and varying the backgate voltage $V_{g}$ from -70V to 70V (in steps of 2V) while measuring the source-drain current $I_{sd}$, and calculating conductance (defined as $I_{sd}/V_{sd}$). Afterwards, we transferred a few-layer GaSe flake onto the device using the PDMS dry-transfer technique and took care to ensure that the GaSe layer does not contact the metal electrodes (Fig.\ref{thresholdshift}(b), bottom image), so that conduction is still dominated by the InSe channel. The heterostructure device is then annealed in an Ar/H\textsubscript{2} environment at 250 \degree C for 1 hour, to improve the hetero-interface, and gate-transfer characteristics was measured again in vacuum. We then compare the gate-transfer curves of the same InSe device before and after GaSe transfer. The device images and experimental results are shown in Fig.\ref{thresholdshift} (Data for an additional sample is plotted in Fig.S1, Supplementary Info) .

From Fig.\ref{thresholdshift}(c), we observe a negative threshold shift of $\Delta V_t \simeq$ 10V following GaSe transfer, indicative of electron transfer from GaSe to InSe. The observed n-type gating behavior is typical of few-layer InSe FETs \cite{Sucharitakul2015}. Although bare GaSe nanoflakes show a p-type gating behavior (Fig.S2, Supplementary Info), the negative threshold shift upon GaSe transfer suggests that the Fermi level E\textsubscript{F} in GaSe is higher than InSe and electrons transfer from the GaSe to InSe layer upon hetero-interface formation, as illustrated in the inset of Fig.\ref{thresholdshift}(c). Such charge transfer is expected to cause a small band bending at the interface which will be discussed later. From the capacitance of the SiO\textsubscript{2} dielectric layer, the charge transfer to the InSe channel is estimated from the observed threshold shift to be $\Delta n = 7 \times 10^{11}$ /cm\textsuperscript{2} (Section 5, Supplementary Info). It is interesting to point out that there is little change in the slope of gate transfer curve in Fig.\ref{thresholdshift}(c) after GaSe transfer, suggesting very little effect of the GaSe layer on the electron field effect mobility of InSe (estimated to be ~500 cm\textsuperscript{2}/Vs from the transconductance\cite{Sucharitakul2015}). Besides revealing the interfacial charge transfer across the GaSe/InSe interface, this experiment also shows that the fabrication process produced high quality interface and there is a good contact between GaSe and InSe. 

\begin{figure*}[t]
\includegraphics[width=16cm]{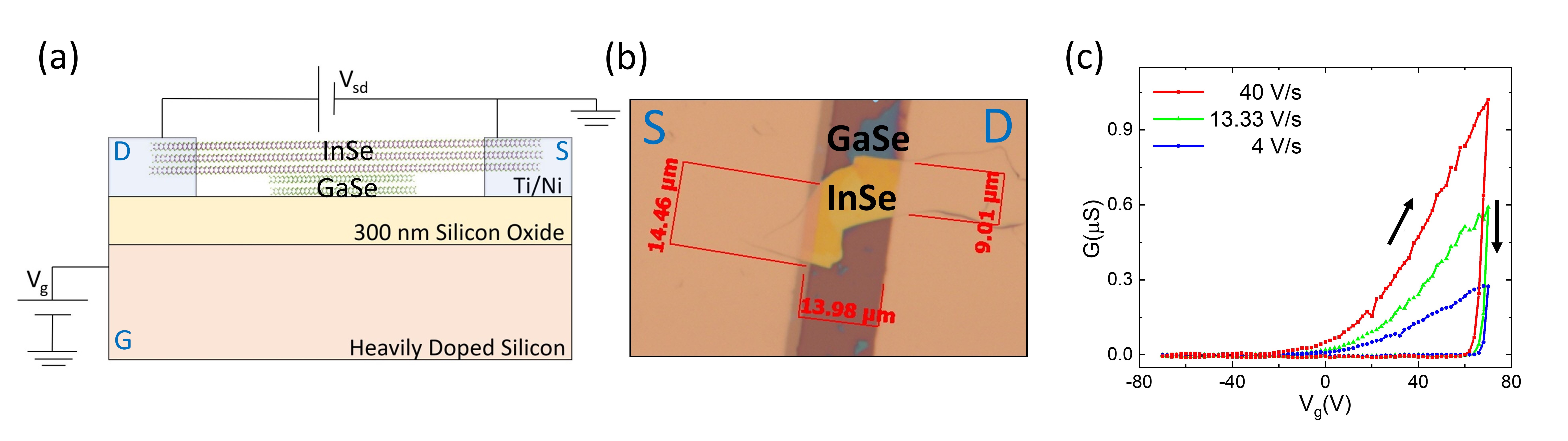}
\caption{\label{fig:epsart} InSe/GaSe heterostructure gating behavior. (a) InSe/GaSe heterostructure device schematic with GaSe layer placed directly on SiO\textsubscript{2} (side view). (b) Device optical image (top view). (c) Gate-transfer curves at different $V_g$ scanning rates and $T$=300K, in vacuum.}
\label{hysteresis}
\end{figure*}

\begin{figure*}[t]
\includegraphics[width=16cm]{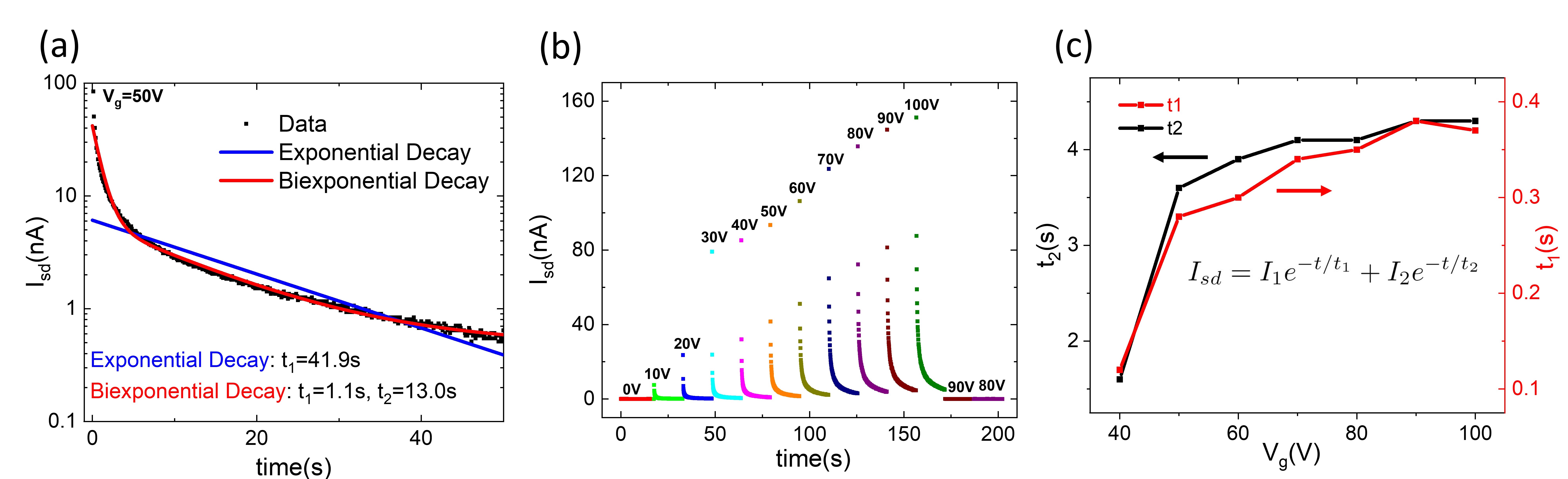}
\caption{\label{fig:epsart} Time-dependent conductance measurements. (a) Time-dependence of conductance upon setting $V_g=50V$ (from $V_g=0V$), for $V_{sd}=0.5V$, plotted in log-linear form and fit to exponential and bi-exponential current decay. (b) Time-dependence of conductance over gate voltage sweep range, with gate-voltage increments of 10V - $V_g$ value specified on top of each curve. (c) Time constants $t_1$ and $t_2$ extracted from bi-exponential fits of data shown in (b).}
\label{time dependance}
\end{figure*}

We next turn to the understanding of InSe/GaSe heterostructures when there is an external gate bias applied across the interface. Fig. \ref{hysteresis} shows a device structure with InSe supported on top of a few-layer GaSe, where the InSe conducting channel is gated through the few-layer GaSe in addition to the SiO\textsubscript{2} dielectric layer. The supporting GaSe layer is not in direct contact with metal electrodes, which prevents direct conduction through GaSe (Fig. \ref{hysteresis} (a)). The gate-transfer curves for this type of devices exhibit n-type behaviour, consistent with the conduction being dominated by n-type InSe (I-V characteristics showing ohmic behavior over the $V_g$ range is plotted in Fig.S3 in Supplementary Info) . In contrast, we observed p-type conduction for bare GaSe (Fig. S2 in Supplementary Info). However, we see that the conductance upon gating for this type of device is two orders of magnitude lower than our bare InSe devices on SiO\textsubscript{2} (as in Fig.2). Further, there is a strong scanning rate dependant hysteresis in the gate-transfer curves. From Fig. \ref{hysteresis} (c) (red curve), we see that, in the direction of increasing gate voltage, the conductance increases slowly from an OFF state ($<$ 10\textsuperscript{-2} \textmu S) to $\sim$ 1 \textmu S when $V_{g}$ changes from -20V to 70V. However, in the direction of decreasing gate voltage, the current decreases quickly from $\sim$ 1 \textmu S to an OFF state within the space of 10V. This indicates a strong hysteresis effect, evidenced by a large shift in threshold voltage of $\Delta V_t \simeq $ 60V between the increasing and decreasing gate voltage scans. A similar hysteretic effect is seen for the different $V_g$ scanning rates shown in Fig. \ref{hysteresis}(c) where,in addition, we observe a strong scanning rate dependence in the gate-transfer curves with peak conductance (i.e conductance at $V_{g}$=70V) reducing from 1 \textmu S to 0.25 \textmu S as the gate voltage scanning rate changes from 40 V/s to 4 V/s. These results indicate a strong time-dependant component in the gating mechanism of these heterostructure devices, which we explore further in this study. Experiments shown in Fig.\ref{hysteresis} and \ref{time dependance} were done in ambient atmosphere, and those in Fig.\ref{temperature} were done in an inert atmosphere of 6-9 Torr of He but no significant difference in the hysteresis features studied here was observed due to the effect of ambient atmosphere when compared to measurements in inert He environment or a vacuum of 50 mTorr.

Fig.\ref{time dependance} shows the results of time-dependant conductance measurements on InSe/GaSe heterostructure devices with the structure shown in Fig.\ref{hysteresis}(a). For these measurements, the source-drain voltage is fixed at 0.5V and a specific constant gate voltage is suddenly switched on and the time-dependence of the source-drain current (and hence the conductance) is measured. For the results shown in Fig. \ref{time dependance}(a), we carry out time dependant measurements after setting $V_g$=50V (from $V_g$=0V). For the results shown in Fig. \ref{time dependance}(b), we measure the time-dependence of conductance after subsequently setting gate voltages in intervals of 10V from 0-100V. 

In Fig. \ref{time dependance}(a), We observe a decay in source-drain current after setting $V_g$=50V and recognize that it has good agreement with a bi-exponential decay model (equation shown in Fig. \ref{time dependance} (c)). In Fig. \ref{time dependance} (b), we observe a similar conductance decay for each gate voltage step in the increasing scanning direction, with good agreement to the bi-exponential decay equation (shown in Fig.\ref{time dependance}(c)) for $V_g>$ 30V. The gate voltage dependence of the two time constants from the bi-exponential fit are shown in Fig.\ref{time dependance}(c). We observe that the time constants of the bi-exponential decay are of the order of 3s and 0.3s respectively, and show a gradual increase with increasing gate voltage. However, when gate voltage is decreased, a drop in conductance to a nearly zero-current OFF state is seen even at $V_g$ = 90V, at a rate that is much faster than the time dependence in the increasing gate voltage measurements. These observations are consistent with the scanning-rate dependant hysteretic gate-transfer curves shown in Fig.\ref{hysteresis}(c). 

\begin{figure*}[t]
\includegraphics[width=16cm]{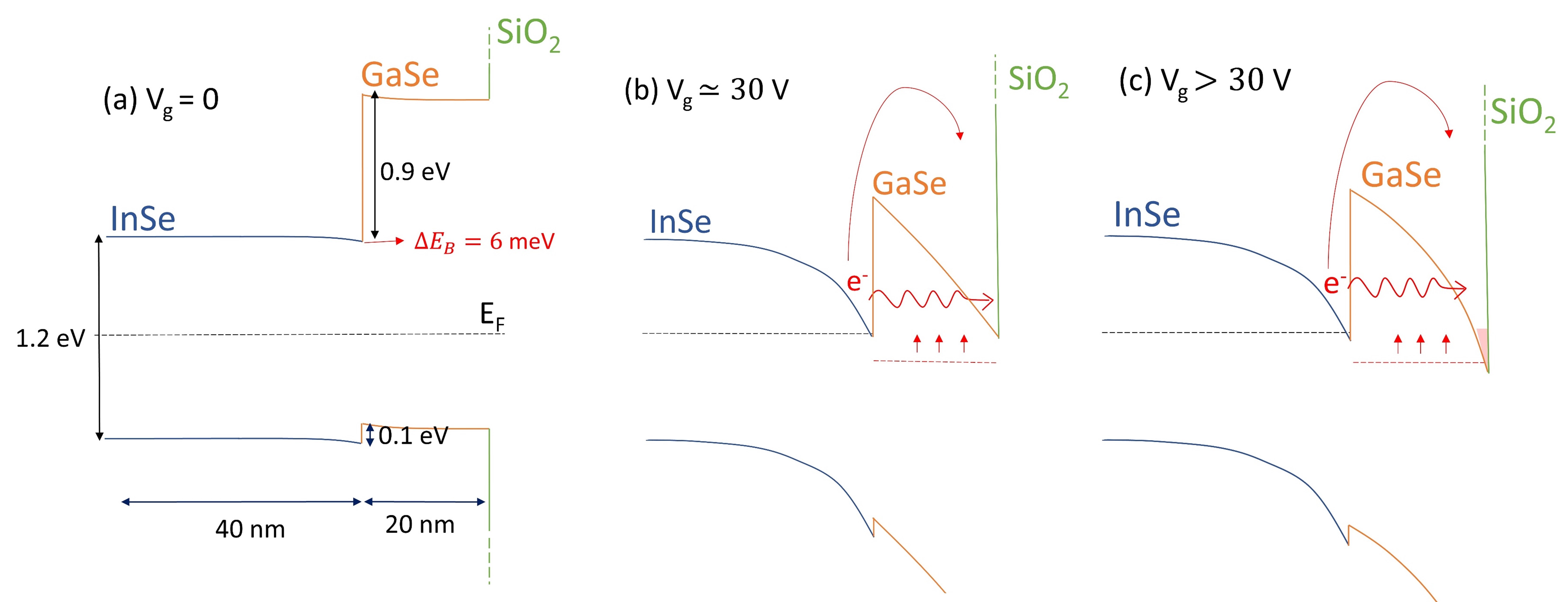}
\caption{\label{fig:epsart} Band diagram for InSe/GaSe heterostructure (shown in Fig.3) at (a) $V_g = 0V$, (b) after applying $V_g \simeq 30V$ and (c) $V_g > 30V$. High $V_g$ induces a strong band tilting in the GaSe layer and significant electron transfer from InSe to GaSe. Shaded region in (c) indicates electrons trapped in GaSe layer due to gate-induced charge transfer from InSe. }
\label{band diagram gating}
\end{figure*}
The hysteresis and time-dependent relaxation behavior suggest there is a mechanism that takes away electrons from the InSe layer when the backgate voltage is increased to transfer electrons from gate to InSe. Such mechanism is likely associated with the electron transferring from InSe to GaSe under an increasing gate bias. This also explains the much faster response of the system when the backgate voltage is decreased to transfer electrons out of InSe, since the electrons flow directly from InSe to gate in such a situation. In previous studies, n-InSe/p-GaSe heterostructures have been shown to form a Type II staggered gap band alignment, with band offsets in the conduction and valence bands determined to be approximately 0.9 eV and 0.1 eV respectively \cite{Balakrishnan2014,Yan2017}. Close to the interface, band bending due to to equilibrium charge transfer will modify this picture, as discussed previously in the experiment shown in Fig.\ref{thresholdshift}. We can estimate the extent of band bending that would result from the previously estimated charge transfer of $\Delta n = 7 \times 10^{11}$ /cm\textsuperscript{2} from GaSe to InSe, by calculating the corresponding change in $E_{F}$ using the 2D density of states. We find that the band-bending at the InSe/GaSe interface to be $\Delta E_{B} = 6$ meV (Section 5, Supplementary Info), which is insignificant compared to the band offsets. The completed band diagram relevant to our InSe/GaSe heterostructure device in Fig.\ref{hysteresis} is therefore shown in Fig.\ref{band diagram gating}(a) (Typical InSe and GaSe nanoflake thicknesses for our heterostructures were estimated to be 40 nm and 20 nm respectively). 

Further, we quantitatively estimate the effect of applying a positive gate voltage through the SiO\textsubscript{2} gate dielectric, on this band diagram. The gate electric field would cause the bands to tilt down, with a slope given by said electric field. At $V_g\simeq 10V$, the conduction band minimum (CBM) of InSe, tilted by electric field, is brought down to the Fermi level $E_F$ at the InSe/GaSe interface ($E_F$ of InSe is estimated to be $\sim$ 0.6 eV below the CBM at $V_g$=0V, see discussion in Section 5, Supplementary Info). The CBM of GaSe is also tilted downward. However, at these low gate-voltages, the CBM of GaSe is still higher than $E_F$ of InSe. Increasing $V_g$ further to $V_{g} \sim 30V$, we estimate a band tilting of 0.9 eV across the 20 nm GaSe layer, which brings the CBM of GaSe at the GaSe/SiO\textsubscript{2} interface to the level of InSe's $E_F$ (Fig. \ref{band diagram gating}(b)) (further details of calculation shown in Section 5, Supplementary Info). Such strong band tilting in the GaSe layer at $V_{g} \gtrsim 30V$ would cause electrons accumulated in InSe to transfer into GaSe and trapped at the GaSe/SiO\textsubscript{2} interface, giving rise to the conductance/current in InSe layer to decay with time, as shown in Fig. \ref{band diagram gating}(b),(c). 


We see, from Fig. \ref{band diagram gating}(b), that the band diagram upon increasing the gate voltage creates a triangular potential barrier seen by electrons accumulated in InSe. There are two pathways for electrons populated in the InSe layer to transfer into the GaSe layer. The first one is through thermionic emission-like process where electrons are thermally excited above the potential barrier and the second process is direct tunneling through the barrier. These processes result in the gate voltage-induced electrons in the conducting InSe channel transferring to GaSe, leading to the observed time dependant decay in conductance after a gate voltage increase, as observed in Fig.\ref{time dependance}.

\begin{figure}[]
\includegraphics[width=8cm]{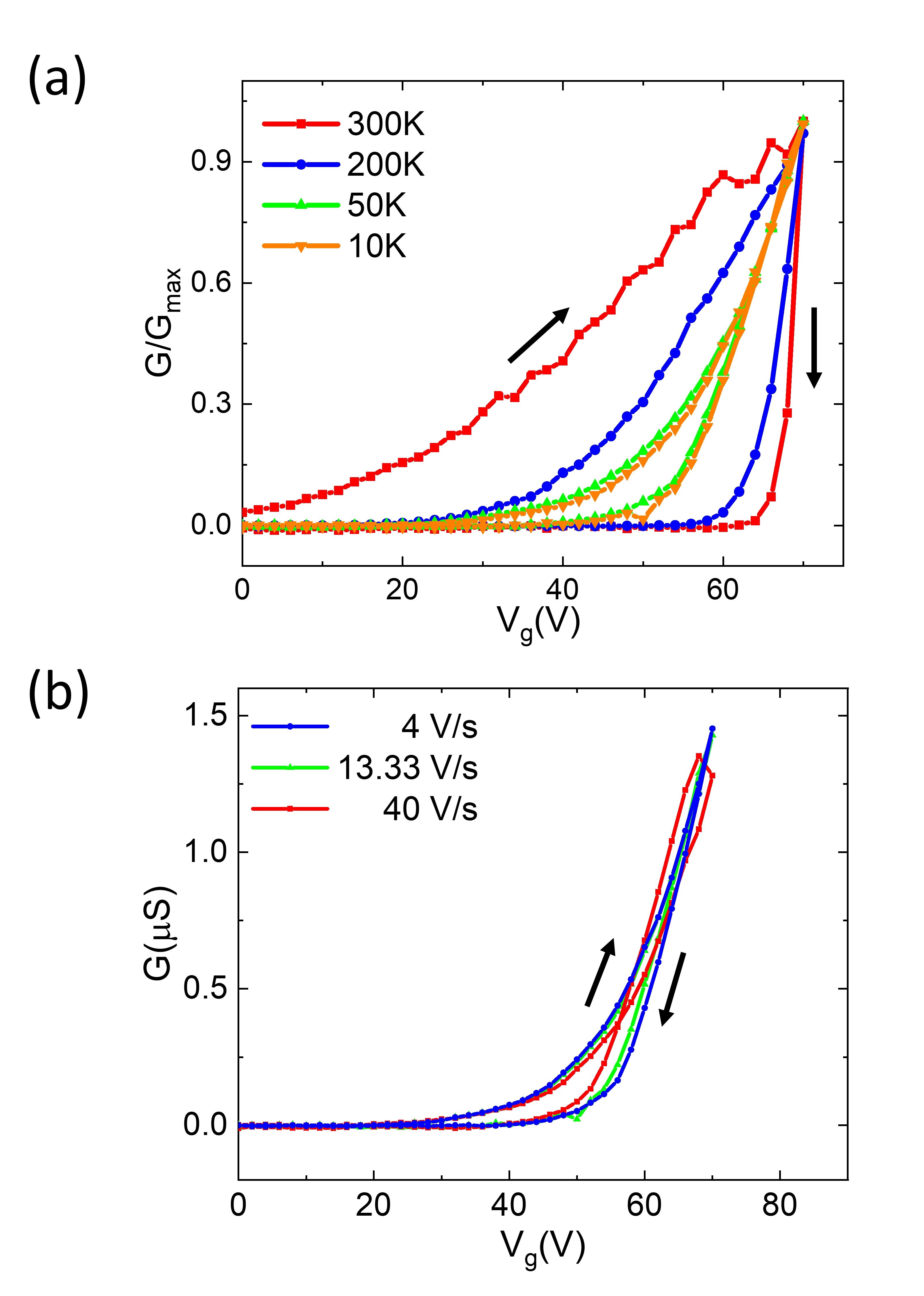}
\caption{\label{fig:epsart} Temperature dependant gating measurements. (a) Normalized gate-transfer curves ($V_g$ scanning rate = 13.33 V/s) plotted at different temperatures. (b) $V_g$ scanning rate dependence of gate-transfer curves at 10K.}
\label{temperature}
\end{figure}

We further consider whether the relaxation time could be associated with the simple RC time constant of charging/decharging the InSe layer as the InSe channel is being depleted (via the electron transferring into GaSe).  If we consider the being depleted InSe channel to have a resistance $R\sim 1 G\Omega$ (from the current $\sim$ 0.5 nA at 0.5 V in Fig.\ref{time dependance}(a)) and capacitance given by $C=C_{g}A$ (where $C_g$ is the gate capacitance per unit area of 300 nm SiO\textsubscript{2} and $A$ is the area of the InSe nanoflake), this  gives us a time constant $\tau_{RC}\sim 10^{-5}$s. This is significantly lesser than the measured time-constants $t_1$ and $t_2$ (Fig. \ref{time dependance}(c)). Therefore, we conclude that the RC charging time of the circuit is not relevant to the observed bi-exponential gate-induced conductance decay. It is therefore likely that this bi-exponential decay is due to the two interfacial charge transfer processes of thermionic emission and electron tunneling across the InSe/GaSe interface. When the gate voltage is decreased, however, these processes do not play a role as the gate-induced electrons are directly pulled from the InSe channel at a much faster rate (set by the RC time constants of the circuit and measurement setup). This results in a rapid conductance decrease over a small decrease in gate voltage, and hence a large hysteresis effect in gate transfer curves, as observed in Fig.\ref{hysteresis}.  
To further investigate the effect of thermal excitation on these observed properties, we perform temperature dependant gate-transfer measurements on these heterostructure devices down to 10K to distinguish processes that are temperature dependent (e.g. thermionic emission) or temperature-independent (e.g. tunneling). We observe, from results shown in Fig.\ref{temperature}(a) which plots the normalized conductance vs. gate voltage at different temperatures for a fixed scanning rate, that the hysteresis effect seen at room temperature (Fig.\ref{hysteresis}) becomes less at lower temperatures and only a weak residual hysteresis is seen below 50K (raw data are shown in Fig.S4 in Supplementary Information). At low temperatures, the scanning-rate of gate voltage also does not have much effect on the gate-transfer measurements as shown in Fig. \ref{temperature}(b) taken at 10K. This indicates that thermionic emission-like thermally assisted charge transfer process no longer plays a role at these low temperatures and any residual hysteresis observed is due to electron tunneling (which also occurs at a much lower rate at lower temperatures - see Hartman et al\cite{Hartman1964}, for example).  

The charge transfer mechanism we have discussed in this paper is in analogy to the one discussed in Li et al \cite{Li2015} to explain large hysteresis in BP/hBN/MoS\textsubscript{2} heterostructures demonstrated for non-volatile memory applications. In Li et al \cite{Li2015}, a 25 nm layer of hBN is used to create the potential barrier across which electrons can tunnel through. In the case of our InSe/GaSe heterostructures, the InSe/GaSe band alignment creates this potential barrier, which is adjusted by the backgate voltage.

\section{Conclusions}
In summary, We have studied electron transport properties of InSe/GaSe heterostructures through DC transport measurements. Under static condition without an external electric field applied across the InSe/GaSe interface, we find evidence of electron transfer from GaSe to InSe layer upon the hetero-interface formation. When an external gate voltage is applied across the InSe/GaSe interface, a bi-exponential conductance decay is observed after setting a gate voltage together with a strong scanning rate dependant hysteresis effect in the gate-transfer curve. We interpret these observations through dynamical interfacial charge transfer between InSe and GaSe, mediated by gate voltage. These observations give insight into transport mechanisms in 2D van der Waal semiconductor heterostructures, and highlight the importance of band alignment in these heterostructure devices. They also open up possibilities for using InSe/GaSe heterostructure's dynamic charge transport properties for device applications (e.g. non-volatile memory).

\begin{acknowledgement}

XPAG acknowledges the financial support from NSF (DMR-1607631) and Raman Sankar and Art Ramirez for providing some of the bulk InSe, GaSe crystals used in this work.
\end{acknowledgement}

\begin{suppinfo}
I-V characterization data for the InSe/GaSe heterostructure sample used in this text (Fig.3), additional data on other samples, gate-transfer curve for a bare GaSe sample, and band bending calculation details are available in Supplementary Info.

\end{suppinfo}

\bibliography{InSeref}

\end{document}